\documentclass[11pt,a4paper]{article}
\usepackage{graphicx,color}
\usepackage{amsmath,amssymb,dsfont,enumerate,fullpage,epsfig,multirow,longtable,bm}
\usepackage{epsfig,epstopdf,hhline}
\usepackage{cases}
\usepackage{algorithm}
\usepackage{algorithmic}

\usepackage[numbers]{natbib}
\usepackage[top=1 in, bottom=1 in, left=1 in, right=1 in, letterpaper]{geometry}
\usepackage{tikz}
\usetikzlibrary{arrows,positioning,decorations.pathreplacing,shapes}

\usepackage{thmtools}
\usepackage{thm-restate}

\newenvironment{proof}{\noindent\emph{Proof\ }}{\hspace*{\fill}$\Box$\medskip}

\newtheorem{theorem}{Theorem}

\newtheorem{lemma}{Lemma}

\newtheorem{corollary}{Corollary}
\newtheorem{remark}{Remark}
\usepackage{amsmath}
\usepackage{paralist}
\usepackage{framed}

\newcommand\restr[2]{{
  \left.\kern-\nulldelimiterspace 
  #1 
  \vphantom{\big|} 
  \right|_{#2} 
  }}

\newcommand{\vect}[1]{\ensuremath{\bm{#1}}}
\newcommand{\one}{\ensuremath{\mathds{1}}}

\newcommand{\E}{\ensuremath{\mathbb{E}}}
\newcommand{\Prob}{\ensuremath{\mathbb{P}}}

\usepackage{mathtools}
\newcommand\givenbase[1][]{\:#1\lvert\:}
\let\given\givenbase

\DeclarePairedDelimiterX\Basics[1](){\let\given\sgiven #1}

\usepackage[utf8]{inputenc} 
\usepackage[T5]{fontenc} 

\usepackage{color, colortbl}
\usepackage{hyperref}
\hypersetup{colorlinks,
            linkcolor=blue,
            citecolor=black,
            urlcolor=magenta,
            linktocpage,
            plainpages=false}
\usepackage[capitalise,noabbrev]{cleveref}

\title{Competitive Algorithms for Online Weighted Bipartite Matching and its Variants}

\author{
Nguyễn Kim Thắng\\
IBISC, University Paris-Saclay, France}

\date{}

\begin{document}

\maketitle

\begin{abstract}
Online bipartite matching has been extensively studied. In the unweighted setting, 
\citet{KarpVazirani90:An-optimal-algorithm} gave an optimal $(1 - 1/e)$-competitive randomized algorithm. In the weighted setting, 
optimal algorithms have been achieved only under assumptions on the edge weights. For the general case,
little was known beyond the trivial $1/2$-competitive greedy algorithm. Recently, 
\citet{FahrbachHuang20:Edge-weighted-online} have presented an 0.5086-competitive algorithm (for the problem in a model, namely free-disposal), 
overcoming the long-standing barrier of 1/2.  
Besides, in designing competitive algorithms for the online matching problem and its variants, several techniques have been developed, in particular 
the primal-dual method. Specifically, 
\citet{DevanurJain13:Randomized-primal-dual} gave a primal-dual framework,
unifying previous approaches and \citet{DevanurJain12:Online-matching} provided another scheme for a generalization of the online matching problem.   

In this paper, we present competitive algorithms for the online weighted bipartite matching 
in different models; in particular we achieve the \emph{optimal} $(1-1/e)$ competitive ratio in the free-disposal model and 
in other model, namely stochastic reward. Our work also unifies the approaches by \citet{DevanurJain13:Randomized-primal-dual} 
and \citet{DevanurJain12:Online-matching} by the mean 
of the primal-dual technique with configuration linear programs.     

\end{abstract}


\section{Introduction}

Matching is fundamental in combinatorial optimization and operations research with wide applications 
from students/colleges admission, kidney exchange to ad-auctions.
Online bipartite matching, motivated by advertising markets, labor markets, etc., has been intensively studied. 
Informally, in the online bipartite matching, there are a set of agents (advertisers) given in advance and 
a set of items (impressions) that are released online one by one. When an item arrives, its edges to the agents are revealed and one 
needs to assign the item irrevocably to an agent (or assign to no one). For unweighted bipartite graphs, 
\citet{KarpVazirani90:An-optimal-algorithm} gave an elegant algorithm \textsc{Ranking} which always outputs a matching of size at least $(1-1/e)$ times that 
of the optimum solution \cite{KarpVazirani90:An-optimal-algorithm,BirnbaumMathieu08:On-line-bipartite}. They also proved that was the best achievable \emph{competitive ratio}. The competitive ratio of an algorithm is defined as  the worst ratio between the objective of the algorithm solution 
and that of the optimum solution. However, for edge-weighted bipartite graphs, no algorithm is competitive with the objective 
of maximizing the total (edge-) weight of the output matching (see for example, \cite{FahrbachHuang20:Edge-weighted-online}).  

In order to circumvent the issue due to edge weights and also to abstract more appropriately 
the practical motivations, in particular the advertising, adwords settings, new models have been proposed. 
In the \emph{free-disposal} model \cite{FeldmanKorula09:Online-ad-assignment}, 
multiple items can be assigned to an agent but only the maximum weight is acknowledged for that agent. 
The objective in this model is to maximize the sum of the heaviest edge weighted assigned to each agent.    
In the \emph{additive-budget} model (Ad-auctions) \cite{MehtaSaberi07:Adwords-and-generalized}, 
each agent has additionally an budget and the revenue received from 
an agent is defined to be the minimum between the total edge weight assigned to the agent and the budget of the agent.
The objective is to maximize the total revenue received from all agents. 
In the \emph{stochastic reward} model \cite{MehtaPanigrahi12:Online-matching}, 
each agent has a weight and each item has additionally a successful probability to be matching to 
an agent. The objective is to maximize the expected number of successfully matched agents, multiplied by their weights.
In the \emph{concave return} model \cite{DevanurJain12:Online-matching}, a generalization of 
the addtive-budget model, the revenue of each an agent is a concave function on the total edge weight assigned to the agent. 
The objective is again to maximize the total revenue received from all agents.  
  
The online matching problem has been extensively studied in those models. 
In the free-disposal model, \citet{FahrbachHuang20:Edge-weighted-online} have recently made a breakthrough by providing a $0.5086$-competitive algorithm,
breaking the long-standing competitive ratio barrier of $1/2$. It revives the hope of potential improvements toward the upper bound of 
$(1-1/e)$ on this problem. 
In the additive-budget model,  $(1-1/e)$-competitive algorithms have been given with additional assumptions: either the edge weights are small
compared to the agents's budgets  \cite{MehtaSaberi07:Adwords-and-generalized,BuchbinderJain07:Online-primal-dual} 
or the weight of each edge is the same for every agent (agent-independent) \cite{AggarwalGoel11:Online-vertex-weighted}. 
In this model, the existence of an $(1-1/e)$-competitive algorithm has been conjectured but still remains open.  
In the concave-return model, \citet{DevanurJain12:Online-matching} gave an optimal online algorithm where the competitive ratio is characterized by a system of differential equations.
    
The primal-dual method has been widely used to study the online bipartite matching problem and its generalizations (including the aforementioned models).     
\citet{DevanurJain13:Randomized-primal-dual} provided an elegant online primal-dual framework that unified previous results by showing how they arose from essentially the same dual update function. 
That unifying technique paves a way for many developments on online matching and its variants. However, there is an exception, another primal-dual 
scheme by \citet{DevanurJain12:Online-matching} in the concave-return model, which is not encompassed by the framework in \cite{DevanurJain13:Randomized-primal-dual}. It is intriguing to understand the nature of those different 
schemes, that potentially leads to improved results for the online matching problems.        


\subsection{Our Contributions}

In this paper, we present an unified primal-dual approach based on configuration linear programs
and give \emph{optimal} algorithms with competitive ratio of $(1-1/e)$ for online matching problem in the free-disposal and the stochastic reward models,
resolving long-standing open questions. Moreover, we provide competitive algorithms for a general model that captures the concave return model 
as a special case. The variable update schemes are built on previous work, in particular \cite{DevanurJain13:Randomized-primal-dual,DevanurHuang16:Whole-page-optimization}. 
The key element is the use of the primal-dual method with configuration LPs
that allows improvements and the unification. 
This can be seen as the last piece to complete the picture which have been widely drawn by previous works.   

In the following, we define a generalization of the online bipartite matching problem, present our approach and the results.

\subsubsection{Model and Approach}		\label{sec:model}

\paragraph{General problem.} We are given a bipartite graph 
$G(L \cup R, E)$ where vertices $L$ on the left-hand side are given in advance and vertices $R$ on the right-hand 
side are released in an online manner. When an online vertex $j \in R$ arrives, its incident edges 
are revealed and an algorithm decides to assign vertex $j$ to an offline neighbour in $L$ or not to assign $j$ to any neighbor. 
The reward function $c: 2^{E} \rightarrow \mathbb{R}_{\geq 0}$ is given so that 
if $M$ is an assignment of online vertices to offline vertices, 
the reward received from this assignment is $c(M)$. Note that if $M$ is a infeasible assignment (an online vertex is assigned to strictly more than one neighbor) 
then one can define $c(M) = 0$.
The objective is to maximize the reward of the output assignment.  

This problem generalizes the models mentioned earlier. Given an assignment $M$, let $M_{i}$ be the set of online vertices assigned to $i$ in $M$.  
In the free-disposal model, an edge between $j \in R$
and $i \in L$ has weight $w_{ij} \geq 0$; several online vertices can be assigned to an offline vertex $i$
but only the weight of the heaviest edge counts in the final matching.
Hence, the reward function can be defined as 
$\sum_{i \in L} \max_{j \in M_{i}} w_{ij}$. In the additive-budget model, each offline vertex $i$ additionally has a budget $W_{i}$
and given an assignment $M$ of online vertices to offline vertices, the reward can be expressed as 
$c(M)= \sum_{i \in L} \min \{W_{i}, \sum_{j \in M_{i}} w_{ij}\}$. In the online matching problem with stochastic rewards, each offline vertex $i$ 
has a weight $w_{i}$ and each online vertex $j$ has a successful matching probability $p_{ij}$ to offline vertex $i$. 
That probability, known in advance, means that if an algorithm decides to match $j$ to $i$, this event successes with probability $p_{ij}$ 
(and with probability $1-p_{ij}$, $i$ and $j$ are not matched).  The goal is to maximize 
the expected number of successfully matched offline vertices, multiplied by their weights. 
In this problem, given an assignment $M$ of online vertices to offline vertices, the reward is
$c(M) = \sum_{i \in L} w_{i} \cdot \min\{1, \sum_{j \in M_{i}} p_{ij}\}$.

\paragraph{Approach.}
The changing point of view in our approach, compared to the previous ones, is to deal directly with the non-linear objective functions. 
Consider the additive-budget model with the reward function $c(M)= \sum_{i \in L} \min \{W_{i}, \sum_{j \in M_{i}} w_{ij}\}$.  
In the previous approaches, the problem is typically formulated as
$$
\max \sum_{i,j} w_{ij} x_{ij} \quad \text{s.t} \quad \sum_{i} x_{ij} \leq 1 ~\forall j, \sum_{j} w_{ij} x_{ij} \leq W_{i}~\forall i, x_{ij} \geq 0~\forall i,j,
$$
in which the reward function is ``linearized'' by using additional constraints (the second constraint). 
In our approach, we directly study the reward function $c(M)= \sum_{i \in L} \min \{W_{i}, \sum_{j \in M_{i}} w_{ij}\}$. The latter is not 
linear and it raises issues to current techniques. In order to circumvent this obstable, 
we consider the primal-dual approach based on configuration LPs \cite{Thang20:Online-Primal-Dual}. The configuration LPs 
have been used by \citet{HuangZhang20:Online-primal} for the stochastic-reward matching problem but their approach is different to ours.

First, we formulate an LP for the general problem. Let $x_{ij}$ be a variable indicating whether $j \in R$ is assigned to $i \in L$.
Let $z_{M}$ be a variable indicating whether an assignment $M \subseteq E$ is selected (output assignment). Consider the following formulation
and the dual of its relaxation.

\begin{minipage}[t]{0.45\textwidth}
\begin{align*}
 & & \max  \sum_{M \subseteq E} & c(M) z_{M} & &\\
(\overline{\alpha}_{j}) & &  \sum_{i \in L} x_{ij}  &\leq 1 & &  \forall j \in R \\
(\overline{\beta}) & & \sum_{M \subseteq E} z_{M} &= 1 & &  \\
(\overline{\gamma}_{i,j}) & & \sum_{M: (i,j) \in M} z_{M} &=  x_{ij} & & \forall i \in L, j \in R \\
& & x_{ij}, z_{M} &\in \{0,1\} & & \forall i \in L, S \subseteq R \\
\end{align*}
\end{minipage}
\quad
\begin{minipage}[t]{0.5\textwidth}
\begin{align*}
\min \sum_{j \in R} \overline{\alpha}_{j} &+ \overline{\beta} & & \\ 
 \overline{\alpha}_{j} &\geq \overline{\gamma}_{i,j} & & \forall i \in L, j \in R \\
\overline{\beta} + \sum_{(i,j) \in M} \overline{\gamma}_{i,j} &\geq c(M) & & \forall M \subseteq E\\ 
\overline{\alpha}_{j} &\geq 0 & & \forall j \in R
\end{align*}
\end{minipage}
In the formulation, the first constraint ensures that an online vertex $j$ can be assigned to at most one online vertex $i$. 
The second constraint guarantees that there must be an output assignment.
The third constraint imposes that if an edge $(i,j)$ is chosen then among all assignments containing $(i,j)$, exactly one will be 
the output assignment. 
 
In the approach, given variables $x_{ij}$ such that $\sum_{i} x_{ij} \leq 1$ for all $j \in R$, 
we always maintain $z_{M} = \prod_{(i,j) \in M} x_{ij} \prod_{(i,j) \notin M} (1 - x_{ij})$ for all $M \subseteq E$. 
By that, the primal constraints $\sum_{M} z_{M} = 1$ 
and $\sum_{i} \sum_{M: (i,j) \in M} z_{M} \leq 1$ always hold true (see Section~\ref{sec:pre} for the detail). 
Let $\vect{x}$ be the vector $(x_{ij})_{(i,j) \in E}$.
The primal objective can be expressed as $C(\vect{x})$ where 
$C: [0,1]^{|E|} \rightarrow \mathbb{R}_{\geq 0}$ is the \emph{multilinear extension} of the reward function $c$, defined as
$$
C(\vect{x}) = \sum_{M \subseteq E} c(M) \prod_{(i,j) \in M} x_{ij} \prod_{(i,j) \notin M} (1 - x_{ij}).
$$
Note that $C(\vect{x})$ can be seen as $\E_{M}[c(M)]$ where an edge $(i,j)$ is randomly included in $M$ with probability $x_{ij}$.  
This view help us to derive updating scheme for dual variables and also make the rounding scheme obvious. 
Given assignment variables $x_{ij}$'s to $i$, it is sufficient to independently round the variables in order to get the objective value of $C(\vect{x})$.

In the primal-dual method, the dual variables often guide the primal assignment via complementary slackness conditions. 
In particular, one complementary slackness condition reads that $x_{ij} > 0$ implies $ \overline{\alpha}_{j} = \overline{\gamma}_{i,j}$.
This indicates the allocation of $j$ to $\arg \max \overline{\gamma}_{i,j}$. That guides the 
strategy of allocating online arrival vertex to offline vertices which are arg max of some terms (corresponding to $\overline{\gamma}_{i,j}$)
in previous algorithms for online matching problems. 
We also adopt this strategy in our algorithms. However, for simplicity and for a better unification/comparaison with previous works, we consider the 
following formulation, which is more compact but equivalent to the previous one (by combining the first and last primal constraints) 
and prove bounds using this formulation. (Even though the aforementioned strategy is less clear from the new dual.)


\begin{minipage}[t]{0.4\textwidth}
\begin{align*}
& & \max  \sum_{M \subseteq E} & c(M) z_{M} & &\\
(\alpha_{j}) \quad & &  \sum_{i \in L} \sum_{M: (i,j) \in M} z_{M}  &\leq 1 & &  \forall j \in R \\
(\beta) \quad & & \sum_{M} z_{M} &= 1 & & \\
& & z_{M} &\in \{0,1\} & & \forall M \subseteq E \\
\end{align*}
\end{minipage}
\quad
\begin{minipage}[t]{0.45\textwidth}
\begin{align*}
\min \sum_{j \in R} \alpha_{j} &+ \beta & & \\ 
\beta + \sum_{i} \sum_{j: (i,j) \in M} \alpha_{j} &\geq c(M) & & \forall M \subseteq E\\ 
\alpha_{j} &\geq 0 & & \forall j \in R
\end{align*}
\end{minipage}
%

\subsubsection{Results}

Building on our approach and the primal-dual schemes of previous works, we provide the following results.
\begin{itemize}
	\item An optimal $(1-1/e)$-competitive algorithm in the free-disposal model.
	\item We revisit the problem in the additive-budget model and give an $(e^{-R_{\max}}-1/e)$-competitive algorithm where 
	$R_{\max} = \max_{i,j} w_{ij}/W_{i}$. This slightly improves the bound of 
	$(1 - R_{\max})\bigr(1 - (1 + R_{\max})^{-1/R_{\max}} \bigr)$ \cite{BuchbinderJain07:Online-primal-dual}.
	More importantly, the algorithm yields the optimal competitive ratio of $(1 - 1/e)$ for the online stochastic-reward matching problem with vanishing probability.
	\item A $(1-\kappa)(1 - 1/e)$-competitive algorithm for the general problem where the reward functions are 
	sub-additive\footnote{A function $c: 2^{E} \rightarrow \mathbb{R}^{+}$ is sub-additive if $c(M_{1} \cup M_{2}) \leq c(M_{1}) + c(M_{2})$ for all $M_{1}, M_{2} \subseteq E$} 
	and $\kappa$ is the curvature of those functions (defined in Section~\ref{sec:sub-additive}).
	\item A competitive fractional algorithm for the general problem with a concavity assumption. 
	We characterize the competitive ratio by system of differential equations. 
	This result recovers the one provided by \citet{DevanurJain12:Online-matching} in the concave-return model.  
\end{itemize}

\subsection{Related works}
There is an extensive literature on online weighted bipartite matching problems. In terms of techniques, significant efforts have been investigated in order to 
unify different approaches. Specifically, \citet{DevanurJain13:Randomized-primal-dual} provided an elegant online primal-dual framework 
showing how different approaches arise from essentially the same dual update function. 
That unifying technique paves a way for many current developments on online matching 
and its variants \cite{DevanurHuang16:Whole-page-optimization,HuangKang20:Fully-online,FahrbachHuang20:Edge-weighted-online,HuangZhang20:Adwords-in-a-Panorama,HuangZhang20:Online-primal}. 
In the following, we summarize the most relevant works to ours and refer the readers to the survey of \cite{Mehta13:Online-Matching}. 

\paragraph{Online matching with free-disposal.}
Having been introduced by \citet{FeldmanKorula09:Online-ad-assignment} in the context of display advertising, the problem 
is widely applied due to its natural economic interpretation \cite{KorulaMirrokni13:Bicriteria-online}. 
However, it had been a long-standing open question whether there exists 
an algorithm with competitive ratio strictly larger than the obvious bound of $1/2$. Recently, in their breakthrough, 
\citet{FahrbachHuang20:Edge-weighted-online} provided 
a 0.5086-competitive algorithm, resolving this question. It revives the hope for improvements towards the upper bound of $1 - 1/e$. 
However, as mentioned in \cite{FahrbachHuang20:Edge-weighted-online}, 
their approach would not lead to a bound better than $5/9$ and in order to obtain a bound closer to $1 -1/e$, 
fundamentally new ideas are required.  

\paragraph{Online matching with additive-budget.}
\citet{MehtaSaberi07:Adwords-and-generalized} introduced the problem (Adwords problem) and gave an optimal $(1-1/e)$ competitive ratio when 
$R_{\max} = \max_{i,j} w_{ij}/W_{i}$ is small. \citet{BuchbinderJain07:Online-primal-dual} simplified the analysis by a primal-dual analysis.
\citet{AggarwalGoel11:Online-vertex-weighted} studied another particular case in which for each $i$, the weights $w_{ij}$'s are the same for every $j$.
They obtained the optimal $(1-1/e)$ competitive ratio with the generalization of the \textsc{Ranking} algorithm \cite{KarpVazirani90:An-optimal-algorithm}.
In this problem (without any assumption), the existence of an $(1-1/e)$-competitive algorithm has been conjectured but still remains open.  
\citet{HuangZhang20:Adwords-in-a-Panorama} recently presented a $0.5016$-competitive algorithm for this problem. 

\paragraph{Online matching with stochastic reward.}
\citet{MehtaPanigrahi12:Online-matching} initiated the study this problem and gave a 0.567-competitive algorithm for the uniform weights and identical 
vanishing probabilities. Moreover, they showed that no algorithm, even in the setting of identical vanishing probabilities,  
has a competitive ratio better than $0.621 < (1 - 1/e)$ against a natural LP. 
Recently, \citet{GoyalUdwani20:Online-Matching} gave an $(1 - 1/e)$-competitive algorithm for the setting 
where the (vanishing) probabilities $p_{ij}$ can be decomposed as $p_{ij} = p_{i} p_{j}$ for 
all $i,j$ (so this includes the identical vanishing probability setting as a particular case).
Independently, \citet{HuangZhang20:Online-primal} provided algorithms with competitive ratios of 0.576 and 0.572 in the settings of vanishing equal probabilities 
and vanishing unequal probabilities, respectively.  

\paragraph{Online matching with concave return.}
\citet{DevanurJain12:Online-matching} considered a generalization of the Adwords problem in which fractional allocation is allowed and 
the rewards are arbitrary monotone concave function. 
They characterized the optimal achievable competitive ratio by system of differential equations and provided matching upper and lower bounds.
The primal-dual scheme by \citet{DevanurJain12:Online-matching} is not captured by the framework of \cite{DevanurJain13:Randomized-primal-dual}. 
It is intriguing to understand the nature
of those algorithms and potentially unify them in a principle approach for online matching problems.

\section{Preliminaries}	\label{sec:pre}
We provide some useful facts and notations. We use bold letters, for example $\vect{x}, \vect{v}$, to denote vectors. 
Let $\mathcal{M}$ be a set of all feasible sub-assignments of online vertices to offline vertices. 
Recall that for any sub-assignment $M \in \mathcal{M}$, each online vertex is assigned to at most one offline vertex.
Given $M \in \mathcal{M}$, denote $M_{i} := \{j \in R: (i,j) \in M\}$ (so $M_{i} \cap M_{i'} = \emptyset$ for any feasible $M \in \mathcal{M}$). 
In the paper, we consider the following formulation and its dual. 

\begin{minipage}[t]{0.45\textwidth}

\begin{align*}
& & {\textsc{Primal}} \\
& & \max  \sum_{M} & c(M) z_{M} & &\\
(\alpha_{j}) \quad & &  \sum_{i \in L} \sum_{M: (i,j) \in M} z_{M}  &\leq 1 & &  \forall j \in R \\
(\beta) \quad & & \sum_{M \in \mathcal{M}} z_{M} &= 1 & & \\
& & z_{M} &\in \{0,1\} & & \forall M \in \mathcal{M} \\
\end{align*}
\end{minipage}
\quad
\begin{minipage}[t]{0.5\textwidth}
\begin{align*}
{\textsc{Dual}} & \\
\min \sum_{j \in R} \alpha_{j} &+ \beta & & \\ 
\beta + \sum_{i \in L} \sum_{j \in M_{i}} \alpha_{j} &\geq c(M) & & \forall M \in \mathcal{M}\\ 
\alpha_{j} &\geq 0 & & \forall j \in R \\
\end{align*}
\end{minipage}
In order to prove the competitive ratio of an algorithm, we will bound the objective value of the primal (due to the decisions of the algorithm) 
and that of the dual (by a dual feasible solution). 

Given a reward function on the assignments $c: \mathcal{M} \rightarrow \mathbb{R}_{\geq 0}$,  
let $C: [0,1]^{|E|} \rightarrow \mathbb{R}_{\geq 0}$ be the multilinear extension of $c$, defined as
$$
C(\vect{x}) = \sum_{M \in \mathcal{M}} c(M) \prod_{(i,j) \in M} x_{ij} \prod_{(i,j) \notin M} (1 - x_{ij}).
$$
where $\vect{x} = (x_{ij})_{(i,j) \in E}$.
$C(\vect{x})$ can be seen as $\E_{M}[c(M)]$ where an edge $(i,j)$ is randomly included in $M$ with probability $x_{ij}$.
The increasing rate of $C(\vect{x})$ while varying $x_{ij}$ is
\begin{align}	\label{eq:increase-C}
\frac{\partial C(\vect{x})}{x_{ij}} 
&= C(\vect{x}_{-(i,j)}, 1) - C(\vect{x}_{-(i,j)}, 0) \notag \\
&= \E_{M' \sim \vect{x}_{-(i,j)}}[c(M' \cup (i,j)) - c(M')] 
\end{align}
where $\vect{x}_{-(i,j)}$ denote the vector $\vect{x}$ without coordinate $(i,j)$. Here,
$M' \sim \vect{x}_{-(i,j)}$ is a random assignment in $\mathcal{M}$ in which an edge $(i',j') \neq (i,j)$ is included with probability $x_{i',j'}$.

Given variables $x_{ij}$ such that $\sum_{i} x_{ij} \leq 1$ for all $j \in R$, 
in our algorithms we always maintain $z_{M} = \prod_{(i,j) \in M} x_{ij} \prod_{(i,j) \notin M} (1 - x_{ij})$.
The variable $z_{M}$ can be interpreted as the probability that the assignment $M$ is selected. 
By that, $\sum_{M \in \mathcal{M}} z_{M} = 1$. Moreover, for every $j \in R$, 
$$
\sum_{i \in L} \sum_{M: (i,j) \in M} z_{M}  = \sum_{i} \biggl( x_{ij}  \cdot \sum_{M': (i, j) \notin M'} z_{M'} \biggr)
\leq \sum_{i} x_{ij} \leq 1.
$$
Hence, given $x_{ij}$'s satisfying $\sum_{i} x_{ij} \leq 1$ for all $j \in R$, the variables $z_{M}$ are feasible. 

In the paper, we use functions $g(y) = \frac{e^{y - 1}}{1 - 1/e}$ and $G(y) = \frac{e^{y-1} - 1/e}{1 - 1/e}$, the primitive integral of $g(y)$.
Note that both $g, G$ are increasing, $G(0) = 0$, $G(1) = 1$, and $1 -G(y) + g(y) = \frac{1}{1 -1/e}$. 

Moreover, in the paper, we denote random variables by capital letters (e.g., $A$) 
and realizations of random assignments by calligraphic letters (e.g., $\mathcal{E}$). 
Given a realization $\mathcal{E}$, we denote $A \given \mathcal{E}$ as the variable $A$ given $\mathcal{E}$.


\section{Online Weighted Matching with Free-Disposal}


In this section, we consider the online weighted matching problem in the free-disposal model in which the reward function 
$c(M) = \sum_{i} \max_{(i,j) \in M} b_{ij}$ for $M \in \mathcal{M}$. For each $i \in L$, define a function $c_{i}: R \rightarrow \mathbb{R}^{+}$ such that
$c_{i}(S) = \max_{j \in S} b_{ij}$ for $S \subseteq R$. By this notation, $c(M) = \sum_{i \in L} c_{i}(M_{i})$.

\paragraph{Algorithm.} 
Let $X_{ij}$ be a $0-1$ random variable indicating whether $j \in R$ is assigned to vertex $i \in L$, 
and $x_{ij} = \Prob[X_{ij} = 1]$. 
Let $Y_{i}(w)$ be a $0-1$ random variable indicating whether at least one online vertex $j$ such that $w_{ij} \geq w$ is assigned to $i \in L$. 
In other words, $\Prob[Y_{i}(w)]$ is the probability that $i$ receives a reward larger than $w$. 

Informally, when an online vertex $j$ arrives, given the realization $\mathcal{E}$ of the random assignement, 
we select vertices $i \in L$ that maximize $w_{i'j} - (B_{i'} \given \mathcal{E})$ among all $i'$ such that $w_{i'j}$ 
is larger than the heaviest weight $w_{i',\sigma(i')}$ currently assigned to $i'$ in $\mathcal{E}$. 
(Note that the selection of $i$ depends on the realization $\mathcal{E}$ of the rounding.)
We continuously increase $x_{ij}$ by $dx$.
By doing that, $\Prob[Y_{i}(w) = 1]$ is increased by $dx$ for $w_{i,\sigma(i)} < w \leq w_{ij}$
(and remain unchanged for other $w$). 
In the algorithm, we also maintain two random variables $A_{j}$ and $B_{i}$ in the algorithm.
We increase variables $A_{j}$ and $B_{i}$ by rules in Step~\ref{algo:FD-dual}.
Intuitively, $A_{j}, B_{i}$ represent the dual variables to be defined later. 
In the end of the while loop when the online vertex $j$ is completely considered, 
$j$ is assigned to $i$ independently with probability $x_{ij}$. 


\begin{algorithm}[ht]
\begin{algorithmic}[1]  
\STATE All primal and dual variables are initially set to 0.
\FOR{each arrival of a new vertex $j$}
	\STATE Let $\mathcal{E}$ be the realization of the random assignment before arrival of $j$.
	\STATE For each $i \in L$, let $\sigma(i)$ be the online vertex with heaviest weight currently assigned to $i$ 
		     in $\mathcal{E}$.
	\WHILE{$\sum_{i} x_{ij} < 1$ \AND $w_{i,\sigma(i)} < w_{ij}$ for some $i \in L$}
		\FOR{every $i \in L$ in $\arg \max_{i'} \{ w_{i',j} - (B_{i'} \given \mathcal{E}): w_{i',j} > w_{i',\sigma(i')}  \}$}	\label{algo:FD-conditions}
			\STATE \label{algo-step:allocation} Increase the probability of assigning $j$ to $i$ by $dx$, i.e., $x_{ij} \gets x_{ij} + dx$.
			\STATE Maintain $Y_{i}(w) = X_{ij}$ for $w_{i,\sigma(i)} < w \leq w_{ij}$. \\
				(The previous step implies $\Prob[Y_{i}(w) = 1] \gets \Prob[Y_{i}(w) = 1] + dx$ for $w_{i,\sigma(i)} < w \leq w_{ij}$.)
			\STATE Increase $B_{i}$ and $A_{j}$ by the following rules respectively,	\label{algo:FD-dual}
			$$
			dB_{i} = \biggl ( \int_{w_{i,\sigma(i)}}^{w_{ij}} g(Y_{i}(w))dw | \mathcal{E} \biggr ) dx
			\qquad \text{and} \qquad
			dA_{j} = \left(w_{ij} - B_{i} \right) dx.
			$$
		\ENDFOR
	\ENDWHILE
	\STATE For every $i$, assign $j$ to $i$ with probability $x_{ij}$.
\ENDFOR
\end{algorithmic}
\caption{Algorithm for Edge-Weighted Matching with Free Disposal.}
\label{algo:free-disposal}
\end{algorithm}

\paragraph{Primal/dual variables.}
Given $x_{ij}$'s, we always maintain primal variables $z_{M} = \prod_{(i,j) \in M} x_{ij} \prod_{(i,j) \notin M} (1 - x_{ij})$
for every $M \in \mathcal{M}$. As argued in Section~\ref{sec:pre}, these primal variables are feasible. 
Moreover, we define dual variables as $\alpha_{j} := \E[A_{j}]$ and $\beta := \sum_{i \in L} \beta_{i}$ where $\beta_{i} := \E[B_{i}]$.

\begin{lemma}		\label{lem:FD-beta}
For any realization $\mathcal{E}$ of the assignment during the execution of the algorithm, the following inequality holds.
$$
\E \bigl [ B_{i} \given \mathcal{E} \bigr] \geq \E \biggl [ \int_{0}^{\infty} G(Y_{i}(w))dw \given \mathcal{E} \biggr]
$$
\end{lemma}
\begin{proof}
We prove by induction on $R$. Initially, when $R = \emptyset$, both sides are 0 so the identity hold trivially.
Assume that the inequality holds for any realization of the random assignment over online vertices released before $j$. 
Let $\mathcal{E}$ be an arbitrary realization of the random assignment before $j$ arrives.  
Conditional on $\mathcal{E}$, for every $i \in L$, we have
\begin{align*}
\E \bigl[ dB_{i}  \given  \mathcal{E} \bigr] = \E \biggl[ \int_{w_{i,\sigma(i)}}^{w_{ij}} g(Y_{i}(w))dw  \given  \mathcal{E} \biggr] dx
= \E \biggl[ \int_{0}^{\infty}  g(Y_{i}(w)) dY_{i}(w)  dw  \given \mathcal{E} \biggr]
= \E \biggl[ \int_{0}^{\infty} dG(Y_{i}(w)) dw \given \mathcal{E} \biggr].
\end{align*}
The second equality holds since, conditional on $\mathcal{E}$, during the increase of $x_{ij}$   
$Y_{i}(w) = 1$ for $w \leq w_{i,\sigma(i)}$; so $dY_{i}(w) = 0$ for $w \leq w_{i,\sigma(i)}$ during this period.
(Recall that if $i$ is not selected in Step~\ref{algo:FD-conditions} then the increase of $x_{ij}$ is 0.)

Integrating both sides of the above equality and note that $G(0) = 0$ and $B_{i} \geq 0$ at any time, we get
$$
\E \bigl[ B_{i} \given \mathcal{E} \bigr] \geq \E \biggl[ \int_{0}^{\infty} G(Y_{i}(w)) dw | \mathcal{E} \biggr]
$$
\end{proof}

\begin{lemma}
The dual solution is feasible, i.e., $\beta + \sum_{i} \sum_{j: j \in M_{i}} \alpha_{j} \geq c(M) ~ \forall M \in \mathcal{M}$.
\end{lemma}
\begin{proof}
Recall that for $M \in \mathcal{M}$, $M_{i} = \{j \in R: (i,j) \in M \}$ and $c_{i}(S) = \max_{j \in S} b_{ij}$ for $S \subseteq R$. 
By definition of the dual variables and the reward function, we need to prove that for every $M \in \mathcal{M}$
$$
\sum_{i \in L} \beta_{i} + \sum_{i \in L} \sum_{j \in M_{i}} \alpha_{j} \geq \sum_{i\in L} \max_{(i,j) \in M} b_{ij} = \sum_{i \in L} c_{i}(M_{i})
$$
Indeed, we will prove a stronger inequality. That is, for any vertex $i \in L$ and subset $S \subseteq R$, it always holds that
$$
\beta_{i} + \sum_{k \in S} \alpha_{k} \geq c_{i}(S)
$$
This inequality subsequently implies the feasibility of the dual variables.  

Fix a vertex $i \in L$ and a subset $S \subseteq R$. 
We prove the above inequality by induction on $R$. 
Initially, when $R = \emptyset$, the inequality holds since both sides are 0. 
Assume that the inequality holds before the arrival 
of vertex $j$.  

\paragraph{Case 1: $j \notin S$ or $j \notin \arg \max_{k \in S} w_{ik}$.} 
Then $c_{i}(S) = c_{i}(S \setminus \{j\})$.   
As the inequality holds before the arrival of $j$, i.e., $\beta_{i} + \sum_{k \in S \setminus \{j\}} \alpha_{k} \geq c_{i}(S \setminus \{j\})$, 
and $\beta_{i}$ is non-decreasing and $\alpha_{j} \geq 0$,
the inequality after the arrival of $j$ also holds, $\beta_{i} + \sum_{k \in S} \alpha_{k} \geq c_{i}(S) = c_{i}(S \setminus \{j\})$.

\paragraph{Case 2: $j \in S$ and $w_{ij} \geq w_{ik} ~\forall j \neq k \in S$ and $\beta_{i} \geq w_{ij}$.} 
The constraint immediately follows again by the non-negativity of $\alpha_{k}$ and $c_{i}(S) = w_{ij}$. 

\paragraph{Case 3: $j \in S$ and $w_{ij} \geq w_{ik} ~\forall j \neq k \in S$ and $w_{ij} > \beta_{i}$.}
We will prove a stronger statement: $\beta_{i} + \alpha_{j} \geq w_{ij}$. 
This will imply $\beta_{i} + \sum_{k \in S} \alpha_{k} \geq \beta_{i} + \alpha_{j} \geq w_{ij} = c_{i}(S)$.

Let $\mathcal{F}$ be the event that $w_{i,\sigma(i)} < w_{ij}$.
In this event, the while loop must have ended with $\sum_{i} x_{ij} = 1$. Throughout the loop (by the condition of the for loop), 
$dA_{j}/dx$ is always at least $\bigl( w_{ij} - B_{i} \bigr)$. Therefore,
$$
(A_{j} \given \mathcal{F})  = \int_{0}^{1} \frac{d(A_{j} \given \mathcal{F}) }{dx}dt 
\geq \int_{0}^{1} \bigl( w_{ij} - (B_{i} \given \mathcal{F}) \bigr) dt = w_{ij} - (B_{i} \given \mathcal{F}).
$$
Hence, $(A_{j} + B_{i} \given \mathcal{F})  \geq w_{ij}$. 

In case of negated event to $\mathcal{F}$, by Lemma~\ref{lem:FD-beta}, we have 
\begin{align*}
\E \biggl [ B_{i} \given \neg \mathcal{F} \biggr] &\geq \E \biggl [ \int_{0}^{\infty} G(Y_{i}(w))dw \given \neg \mathcal{F} \biggr]
\geq  \E \biggl [ \int_{0}^{w_{i,\sigma(i)}} G(Y_{i}(w))dw \biggr] \\
&= \int_{0}^{w_{i,\sigma(i)}} G(1)dw = w_{i,\sigma(i)} \geq w_{ij}
\end{align*}
Combining both cases, we deduce $\beta_{i} + \alpha_{j} = \E[A_{j} + B_{i}] \geq w_{ij}$.
\end{proof}

\begin{theorem}
The randomized Algorithm~\ref{algo:free-disposal} is $(1-1/e)$-competitive.
\end{theorem}
\begin{proof}
We prove that for every online vertex $j$, the increase of the primal is at least $(1 -1/e)$ that of the dual.
Fix an arbitrary realization $\mathcal{E}$ before the arrival of vertex $j$. 
When $x_{ij}$ increases, by Equation~(\ref{eq:increase-C}), the increasing rate of the primal is
$$ 
\E_{M' \sim \vect{x}_{-(i,j)}}[c(M' \cup (i,j)) - c(M') \given \mathcal{E}] = w_{ij} - w_{i,\sigma(i)} 
$$
where $w_{i,\sigma(i)}$ is the heaviest weight assigned to $i$ in $\mathcal{E}$. 

Besides, given a realization $\mathcal{E}$, $\int_{0}^{w_{i,\sigma(i)}} G(Y_{i}(w))dw = \int_{0}^{w_{i,\sigma(i)}} G(1)dw = \int_{0}^{w_{i,\sigma(i)}} 1 dw$.
Therefore, conditional on the realization $\mathcal{E}$,
\begin{align*}
\E \biggl[  \frac{dA_{j}}{dx} \given \mathcal{E} \biggr] &= w_{ij} - \E \biggl[ B_{i} \given \mathcal{E} \biggr]
\leq w_{ij} - \E \biggl[ \int_{0}^{\infty} G(Y_{i}(w)) dw | \mathcal{E} \biggr] \\
&=  \int_{0}^{w_{ij}} 1 dw - \E \biggl[ \int_{0}^{w_{i,\sigma(i)}} G(1) dw \biggr] - \E \biggl[ \int_{w_{i,\sigma(i)}}^{\infty} G(Y_{i}(w)) dw | \mathcal{E} \biggr] \\
&= \E \biggl[  \int_{w_{i,\sigma(i)}}^{w_{ij}} (1 - G(Y_{i}(w))) dw - \int_{w_{ij}}^{\infty} G(Y_{i}(w)) dw \given \mathcal{E} \biggr] \\
&\leq \E \biggl[ \int_{w_{i,\sigma(i)}}^{w_{ij}} (1 - G(Y_{i}(w))) dw \given \mathcal{E} \biggr] 
\end{align*}
where the first inequality is due to Lemma~\ref{lem:FD-beta}.

We deduce that, conditional on the realization $\mathcal{E}$,
\begin{align*}
\E \biggl[ \frac{dA_{j}}{dx}  + \frac{dB_{i}}{dx} \given \mathcal{E} \biggr]
&\leq \E \biggl[  \int_{w_{i,\sigma(i)}}^{w_{ij}} \bigl( 1 - G(Y_{i}(w)) + g(Y_{i}(w)) \bigr) dw  \given \mathcal{E} \biggr] \\
&= \int_{w_{i,\sigma(i)}}^{w_{ij}} \frac{dw}{1-1/e} = \frac{w_{ij} - w_{i,\sigma(i)}}{1 - 1/e}
\end{align*}
Note that whenever $x_{ij}$ increases, only $\beta_{i}$ increases whereas $\beta_{i'}$'s for $i' \neq i$ remain unchanged.
Hence, the increase of the primal is at least $(1 -1/e)$ that of the dual for any realization $\mathcal{E}$.
As it holds for any realization, it holds in expectation and the theorem follows.
\end{proof}


\begin{remark}
\normalfont 
Algorithm~\ref{algo:free-disposal} has the structure similar to the algorithm for the fractional 
version of the online matching with free-disposal \cite[Algorithm 2]{DevanurHuang16:Whole-page-optimization}. 
However, there is a small but crucial difference. 
In \cite[Algorithm 2]{DevanurHuang16:Whole-page-optimization}, 
the update rate of $\beta$-variable is $\int_{w_{i,j'}}^{w_{i,j}} g(y_{i}(w))dw$ where $w_{i,j'}$ is the \emph{smallest}
edge-weight among the weights already assigned to $i$ with positive probability. 
This update is based on the complementary cumulative distribution function viewpoint. 
Our update in Step~\ref{algo:FD-dual} (on the random variable counterparts) 
is guided by the configuration LP approach. Specifically, we update $B_{i}$ at the rate of 
$\int_{w_{i,\sigma(i)}}^{w_{ij}} g(Y_{i}(w))dw$ where $w_{i,\sigma(i)}$ is the \emph{heaviest}
edge-weight among the weights already assigned to $i$ (in the current realization). 
\end{remark}

\begin{remark}
\normalfont 
As a sanity check, we consider Algorithm~\ref{algo:free-disposal} in the online unweighted matching problem, especially applying to 
the graph corresponding to the upper triangular matrix --- it is the worst example for the KVV algorithm of \citet{KarpVazirani90:An-optimal-algorithm}. 
In this setting, the
graph has vertex sets ($L = R = \{1, \ldots, n\}$). One samples a uniformly random permutation $\pi$ of the set $[n]$ and define the edge set
of the graph to be $E = \{(\pi(i), j): i \geq j\}$. Note that $\pi$ is unknown to the algorithm. The online vertices in $R$ arrive in the order $j = 1, 2, \ldots, n$. 
The optimal solution is the perfect matching consisting of the edges $(\pi(j),j)$ for $1 \leq j \leq n$. On this input, at the arrival of an online vertex $j$,
Algorithm~\ref{algo:free-disposal} increases $x_{ij}$ uniformly for all unmatched offline neighbors of $j$. This results in these offline nodes’ dual variables increasing uniformly, resulting in the next allocation of the next iteration again being uniform among the unmatched neighbors, and so on. 
The integral (random) allocation then matches each online node $j$ to each of its free neighbors uniformly random.
Implementing Algorithm~\ref{algo:free-disposal} and comparing to KVV algorithm on this example, we observe in \cite{DurrThang21:Implementation-on-upper} that 
they both have the theoretically predicted competitive ratio ($1 - 1/e \approx 0.632$).
\end{remark}



\section{Online Weighted Matching with Additive Budgets}

In this section, we consider first the online weighted matching problem in the additive-budget model 
in which for $M \in \mathcal{M}$, the reward function $c(M) = \sum_{i \in L} c_{i}(M_{i})$ where $c_{i}(S) = \min \{ \sum_{j \in S} w_{ij}, W_{i} \}$
for $S \subseteq R$, for all $i$. Subsequently, we deduce the performance guarantee for the problem in 
the stochastic-reward model as a corollary.  

\paragraph{Algorithm.} 
Recall $X_{ij}$ is the 0-1 random variable indicating whether $j$ is assigned to $i$.
In the algorithm, we maintain a random variable $Y_{i}$ intuitively (but not exactly) representing the fraction of the consumed budget of $i$.
Moreover, as the previous algorithm, we also maintain additional random variables $A_{j}$ and $B_{i}$ that help the algorithm's decisions 
and the definitions of dual variables. 

Informally, when an online vertex $j$ arrives, given the current assignment $\mathcal{E}$, we continuously increase 
$x_{ij}$ by $dx$ if $i$ maximizes the term $\min \bigl \{ w_{i',j}, \max \bigl \{ 0, W_{i'} - \sum_{j': (i',j') \in \mathcal{E}} w_{i',j'} \bigr \} \bigr \} \cdot \bigl( 1 - \frac{(B_{i'} \given \mathcal{E})}{W_{i'}} \bigr)$ among all $i'$. Here, the coefficient of $\min \bigl \{ w_{i,j}, \max \bigl \{ 0, W_{i} - \sum_{j': (i,j') \in \mathcal{E}} w_{i,j'} \bigr \} \bigr \}$,
as we will argue later, is the rate of the primal when $x_{ij}$ varies. 
(The choice of this coefficient shows the usefulness of the configuration LP approach.)
Intuitively, given an assignment $\mathcal{E}$ before $j$ is released, 
$\min \bigl \{ w_{i,j}, \max \bigl \{ 0, W_{i} - \sum_{j': (i,j') \in \mathcal{E}} w_{i,j'} \bigr \} \bigr \}$ is the increase of the total reward if $j$ is (integrally) assigned to $i$.
Subsequently, we increase variables $A_{j}$ and $B_{i}$ by rules in Step~\ref{algo:AB-dual}.

\begin{algorithm}[ht]
\begin{algorithmic}[1]  
\STATE Initially, all primal/dual variables and the corresponding probabilities of random variables $X_{ij}, Y_{i}, A_{j}, B_{i}$ are set to 0.
\FOR{each arrival of a new vertex $j$}
	\STATE Let $\mathcal{E}$ be the realization of the random assignment before arrival of $j$.
	\WHILE{$\sum_{i} x_{ij} < 1$ and $(Y_{i} \given \mathcal{E}) < 1$ for some $i \in L$}
		\FOR{every $i \in L$ in $\arg \max_{i'} \bigl \{ \min \bigl \{ w_{i',j}, \max \bigl \{ 0, W_{i'} - \sum_{j': (i',j') \in \mathcal{E}} w_{i',j'} \bigr \} \bigr \} \cdot \bigl( 1 - \frac{(B_{i'} \given \mathcal{E})}{W_{i}} \bigr) \bigr \}$}
			\STATE \label{add-step:allocation} Increase the probability of assigning $j$ to $i$ by $dx$, i.e., $x_{ij} \gets x_{ij} + dx$. 
			\STATE Update $Y_{i}$, $B_{i}$ and $A_{j}$ by the following rules respectively,		\label{algo:AB-dual}
			\begin{align*}
			Y_{i} &= (Y_{i} \given \mathcal{E} ) + \min \bigl \{ w_{ij}, W_{i} - \sum_{j': (i,j') \in \mathcal{E}} w_{ij'} \bigr \} \cdot \frac{1}{W_{i}} \cdot X_{ij}, \\
			dB_{i} &= \min \bigl \{ w_{ij}, \max \bigl \{ 0, W_{i} - \sum_{j': (i,j') \in \mathcal{E}} w_{ij'} \bigr \} \bigr \} \cdot g(Y_{i}) dx, \\
			dA_{j} &= \min \bigl \{ w_{ij}, \max \bigl \{ 0, W_{i} - \sum_{j': (i,j') \in \mathcal{E}} w_{ij'} \bigr \} \bigr \} \cdot \left(1 - \frac{B_{i}}{W_{i}} \right) dx.
			\end{align*}
		\ENDFOR
	\ENDWHILE
	\STATE For every $i$, assign $j$ to $i$ with probability $x_{ij}$.
\ENDFOR
\end{algorithmic}
\caption{Algorithm for Edge-Weighted Matching with Additive Budgets.}
\label{algo:additive}
\end{algorithm}

\paragraph{Primal/dual variables.}
As in the previous section, given $x_{ij}$'s, we define primal variables $z_{i,S} = \prod_{j \in S} x_{ij} \prod_{j \notin S} (1 - x_{ij})$
for every $i \in L$, $S \subseteq R$ and dual variables $\alpha_{j} = \E[A_{j}]$, $\beta_{i} = \E[B_{i}]$ and $\beta = \sum_{i\in L} \beta_{i}$.

\begin{lemma}
For any realization $\mathcal{E}$ of the random assignment and for every $i \in L$,
the following invariant always holds
$$
\E [ B_{i} \given \mathcal{E}] =  \E[ W_{i} \cdot G(Y_{i}) \given \mathcal{E}].
$$
\end{lemma}
\begin{proof}
Fix an arbitrary vertex $i \in L$. 
Again, we prove by induction on $R$. For the base case where $R = \emptyset$, both side
are 0 so the invariant holds trivially. Assume that the invariant holds for any realization of the assignment before the arrival of vertex $j$. 
Let $\mathcal{E}$ be a realization of the random assignment before the arrival of $j$. 
During the consideration of $j$, it holds that
\begin{align*}
\E [ dB_{i} \given \mathcal{E}] = \E [ \min \bigl \{ w_{ij}, \max \bigl \{ 0, W_{i} - \sum_{j': (i,j') \in \mathcal{E}} w_{ij'} \bigr \} \bigr \} \cdot g(Y_{i}) dx \given \mathcal{E}]
= \E [ W_{i} g(Y_{i}) dY_{i} \given \mathcal{E}]
\end{align*}
where the last equality is due to the definition of $Y_{i}$.
Integrating both sides, the lemma follows.
\end{proof}

Recall that $R_{\max} = \max_{i,j} w_{i,j}/W_{i}$. The following lemma shows that 
the dual constraints are feasible up to a factor of $G(1 - R_{\max})$.

\begin{lemma}		\label{lem:sub-add-feasible}
For every vertex $i \in L$ and subset $S \subseteq R$, it holds that
$$
\beta_{i} + \sum_{j \in S} \alpha_{j} \geq G(1 - R_{\max}) \cdot c_{i}(S).
$$
Consequently, it implies that for any $M \in \mathcal{M}$,
$
\beta + \sum_{i\in L} \sum_{j \in M_{i}} \alpha_{j} \geq G(1 - R_{\max}) \cdot c(M).
$
\end{lemma} 
\begin{proof}
Fix a vertex $i \in L$ and a subset $S \subseteq R$. 
We prove that, given an arbitrary realization $\mathcal{E}$ of the random assignment, 
\begin{align}	\label{eq:sub-add-feasible-var}
\E \bigl[ B_{i} + \sum_{j \in S} A_{j} \given \mathcal{E} \bigr] \geq G(1 - R_{\max}) \cdot c_{i}(S)
\end{align}

If $(Y_{i} \given \mathcal{E}) = 1$ then $(B_{i} \given \mathcal{E}) = W_{i}$, so the lemma inequality holds. 
In the following, assume that $(Y_{i} \given \mathcal{E}) < 1$ after the arrival of the last vertex in $S$. 
By the while loop condition, $\sum_{i} x_{ij} = 1$ for every $j \in S$. 
For every vertex $j \in S$, let $\mathcal{E}_{j}$ be the assignment before the arrival of $j$. (Note that $\mathcal{E}_{j} \subseteq \mathcal{E}$.)
By the condition of the for loop,  
$dA_{j}/dx$ must be at least $\min \bigl \{ w_{ij}, \max \bigl \{ 0, W_{i} - \sum_{j': (i,j') \in \mathcal{E}_{j}} w_{ij'} \bigr \} \bigr \} \cdot \bigl( 1 - \frac{(B_{i} \given \mathcal{E}_{j})}{W_{i}} \bigr)$. Therefore, given $\mathcal{E}$ (in particular $\mathcal{E}_{j}$),
\begin{align*}
A_{j} = \int_{0}^{1} \frac{dA_{j}}{dx}dt
&\geq \int_{0}^{1} \min \bigl \{ w_{ij}, \max \bigl \{ 0, W_{i} - \sum_{j': (i,j') \in \mathcal{E}_{j}} w_{ij'} \bigr \} \bigr \} 
\cdot \bigl( 1 - \frac{(B_{i} \given \mathcal{E}_{j})}{W_{i}} \bigr) dt \\
&= \min \bigl \{ w_{ij}, \max \bigl \{ 0, W_{i} - \sum_{j': (i,j') \in \mathcal{E}_{j}} w_{ij'} \bigr \} \bigr \} \cdot \bigl( 1 - \frac{(B_{i} \given \mathcal{E}_{j})}{W_{i}} \bigr) \\
&\geq \min \bigl \{ w_{ij}, \max \bigl \{ 0, W_{i} - \sum_{j': (i,j') \in \mathcal{E}_{j}} w_{ij'} \bigr \} \bigr \} \cdot \bigl( 1 - \frac{(B_{i} \given \mathcal{E})}{W_{i}} \bigr).
\end{align*}
where the last inequality holds since the value of $B_{i}$ is non-decreasing as long as online vertices arrive. 

Assume that there exists $k \in S$ such that 
$\min \bigl \{ w_{ik}, \max \bigl \{ 0, W_{i} - \sum_{j': (i,j') \in \mathcal{E}_{k}} w_{ij'} \bigr \} \bigr \} = W_{i} - \sum_{j': (i,j') \in \mathcal{E}_{k}} w_{ij'}$,
meaning that $(Y_{i} \given \mathcal{E}_{k}) \geq 1 - \frac{w_{ik}}{W_{i}}$. In this case, 
\begin{align*}
\E \bigl[ B_{i} + \sum_{j \in S} A_{j} \given \mathcal{E} \bigr]
&\geq \E \bigl[ B_{i} \given \mathcal{E}_{k} \bigr] 
	\geq \E \bigl[ W_{i} G(Y_{i}) \given \mathcal{E}_{k} \bigr] \\
&\geq W_{i} G\biggl( 1 - \frac{w_{ik}}{W_{i}} \biggr)
\geq W_{i} G( 1 - R_{\max}).
\end{align*}

In the remaining, assume that for every $j \in S$, 
$\min \bigl \{ w_{ij}, \max \bigl \{ 0, W_{i} - \sum_{j': (i,j') \in \mathcal{E}_{j}} w_{ij'} \bigr \} \bigr \} = w_{ij}$. 
Given $\mathcal{E}$, we have
\begin{align*}
\E \biggl [ B_{i} + \sum_{j \in S} A_{j} \given \mathcal{E} \biggr]
&\geq \E \biggl [ B_{i} + \sum_{j \in S} w_{ij} \biggl( 1 - \frac{B_{i}}{W_{i}} \biggr) \given \mathcal{E} \biggr] \\
&= \E \biggl [ B_{i} + \biggl( 1 - \frac{B_{i}}{W_{i}} \biggr) \sum_{j \in S} w_{ij} \given \mathcal{E} \biggr] \\
&\geq \E \biggl [ B_{i} + \biggl( 1 - \frac{B_{i}}{W_{i}} \biggr)  \cdot \min \biggl \{ W_{i},  \sum_{j \in S} w_{ij} \biggl \}  \given \mathcal{E} \biggr] \\
&= \min \biggl \{ W_{i},  \sum_{j \in S} w_{ij} \biggl \} + \biggl( 1 - \frac{\min \bigl \{ W_{i},  \sum_{j \in S} w_{ij} \bigl \}}{W_{i}} \biggr) \E \biggl[ B_{i}  \given \mathcal{E} \biggr] \\
&\geq \min \biggl \{ W_{i},  \sum_{j \in S} w_{ij} \biggl \} = c_{i}(S).
\end{align*}
where the second inequality is due to $1 - \frac{W_{i}}{B_{i}} \geq 0$.
As Inequality~(\ref{eq:sub-add-feasible-var}) holds for any realization $\mathcal{E}$, we deduce that 
$$
\beta_{i} + \sum_{j \in S} \alpha_{j} \geq G(1 - R_{\max}) \cdot c_{i}(S).
$$

Consequently, for any $M \in \mathcal{M}$, applying the above inequality for $S = M_{i}$ and summing over all $i \in L$, we have 
$
\beta + \sum_{i\in L} \sum_{j \in M_{i}} \alpha_{j} \geq G(1 - R_{\max}) \cdot c(M).
$
\end{proof}

\begin{theorem}		\label{thm:additive}
The randomized Algorithm~\ref{algo:additive} has the competitive ratio of $G(1 - R_{\max}) \cdot \bigl( 1 - 1/e \bigr) = e^{-R_{\max}} - e^{-1}$.
\end{theorem}
\begin{proof}
We prove that for every online vertex, the increase of the primal is at least $(1 -1/e)$ that of the dual.
Fix an arbitrary realization $\mathcal{E}$ before the arrival of vertex $j$. 
When $dx$ amount of $j$ is allocated to $i$, by Equation~(\ref{eq:increase-C}), the increasing rate of the primal is
$$ 
\E_{M' \sim \vect{x}_{-(i,j)}}[c(M' \cup (i,j)) - c(M') \given \mathcal{E}] = \min \bigl \{ w_{ij}, \max \bigl \{ 0, W_{i} - \sum_{j': (i,j') \in \mathcal{E}} w_{ij'} \bigr \} \bigr \}.
$$

Conditional on the realization $\mathcal{E}$, the expected increasing rate of the dual is 
\begin{align*}
\E \biggl [ \frac{dA_{j}}{dx} + \frac{dB_{i}}{dx} \given \mathcal{E} \biggr] 
&= \min \bigl \{ w_{ij}, \max \bigl \{ 0, W_{i} - \sum_{j': (i,j') \in \mathcal{E}} w_{ij'} \bigr \} \bigr \} \cdot \E \left[ 1 - \frac{B_{i}}{W_{i}}  + g(Y_{i}) \given \mathcal{E} \right] \\
&= \min \bigl \{ w_{ij}, \max \bigl \{ 0, W_{i} - \sum_{j': (i,j') \in \mathcal{E}} w_{ij'} \bigr \} \bigr \} \cdot \E \left[ 1 - G(Y_{i})  + g(Y_{i}) \given \mathcal{E} \right] \\
&= \min \bigl \{ w_{ij}, \max \bigl \{ 0, W_{i} - \sum_{j': (i,j') \in \mathcal{E}} w_{ij'} \bigr \} \bigr \} \cdot \frac{1}{1 - 1/e}.
\end{align*}
So, given a realization $\mathcal{E}$, the ratio between the increasing rates of the primal and the dual is $1-1/e$. 
This holds for any realization. Therefore, by Lemma~\ref{lem:sub-add-feasible} and the weak duality, the competitive ratio 
of Algorithm~\ref{algo:additive} is $G(1 - R_{\max}) \cdot \bigl( 1 - 1/e \bigr) = e^{-R_{\max}} - e^{-1}$.
\end{proof}


\begin{corollary}
Algorithm~\ref{algo:additive} is $(1-1/e)$-competitive for the online stochastic-reward matching problem
with vanishing probability.
\end{corollary}
\begin{proof}
In the online matching problem with stochastic rewards, the reward function $c(M)$ can be expressed as $\sum_{i \in L} c_{i}(M_{i})$
where $c_{i}(S) = w_{i} \cdot \min\{1, \sum_{j \in S} p_{ij}\} = \min\{w_{i}, \sum_{j \in S} p_{ij}w_{i}\}$ for $S \subseteq R$. 
Applying Theorem~\ref{thm:additive} and using the vanishing probability property ($R_{\max} = \max p_{ij} \rightarrow 0$), 
we deduce the competitive ratio of $(1-1/e)$.
\end{proof}




\section{Online Weighted Matching with Sub-Additive Rewards}	\label{sec:sub-additive}
In this section, we consider the online weighted matching problem in a general model 
in which the reward function is sub-additive, i.e., $c(M_{1} \cup M_{2}) \leq c(M_{1}) + c(M_{2})$.
Without loss of generality, assume that $c(M) \leq 1$ for all $M$ (this can be done by scaling).

Given a sub-additive function $f: 2^{E} \rightarrow [0,1]$, 
define the \emph{total curvature} $\kappa_{f}$ of $f$ as 
\begin{align}	\label{def:curvature}
\kappa_{f} = 1 - \min_{\emptyset \neq M \subseteq E} \min_{e \in M} \frac{f(M) - f(M \setminus \{e\})}{f(\{e\})}.
\end{align}
This definition generalizes the notion of curvature for submodular functions introduced by \citet{ConfortiCornuejols84:Submodular-set-functions}.
A function $f: 2^{E} \rightarrow \mathbb{R}_{\geq 0}$ is submodular if $f(M_{1} \cup \{e\}) - f(M_{1}) \geq f(M_{2} \cup \{e\}) - f(M_{2})$ for all 
$M_{1} \subseteq M_{2} \subseteq E$. 
In the context of submodular functions, the curvature \cite{ConfortiCornuejols84:Submodular-set-functions} is defined as 
$1 - \min_{e \in E} \frac{f(E) - f(E \setminus \{e\})}{f(\{e\})}$. The latter is exactly the same as (\ref{def:curvature}) since for submodular functions, 
$\min_{M \subseteq E} \bigl\{ f(M) - f(M \setminus \{e\}) \bigr\} = f(E) - f(E \setminus \{e\})$. 
Intuitively, the total curvature mesures how far away $f$ is from being \emph{modular}. The concept of 
curvature is widely used in the context of submodular optimization; for exemple in determining both upper and lower bounds on the approximation ratios
for many submodular and learning problems \cite{ConfortiCornuejols84:Submodular-set-functions,GoemansHarvey09:Approximating-submodular,BalcanHarvey12:Learning-Submodular,Vondrak10:Submodularity-and-Curvature:,IyerJegelka13:Curvature-and-optimal,SviridenkoVondrak17:Optimal-approximation}. 

Denote $\kappa = \kappa_{c}$. We will bound the competitive ratio in this section as a function of $\kappa$.
Note that if the reward function can be decomposed as $c(M) = \sum_{i \in L} c_{i}(M_{i})$ for every $M \in \mathcal{M}$
where $c_{i}: 2^{R} \rightarrow \mathbb{R}^{+}$ then $\kappa = \kappa_{c} = \max_{i} \kappa_{c_{i}}$.

\paragraph{Algorithm.}
Recall that $C: [0,1]^{|E|} \rightarrow [0,1]$ be the multilinear extension of $c: 2^{E} \rightarrow [0,1]$. 
The algorithm for the online matching with sub-additive reward is 
a generalization of the previous algorithms. The main difference is that in this algorithm, 
we deal directly with fractional variables instead of random variables and realizations of random assignments.  
The reason is that in the previous sections, our goal is to achieve the tight competitive ratio of $(1-1/e)$ 
whereas in this general setting, we aim for a weaker guarantee, given the 
hardness result\footnote{No randomized algorithm is 1/2-competitive unless NP= RP, even for submodular rewards (a sub-class of sub-additive rewards).}
of \cite{KapralovPost13:Online-submodular}.

\begin{algorithm}[ht]
\begin{algorithmic}[1]  
\STATE All primal and dual variables are initially set to 0.
\FOR{each arrival of a new vertex $j$}
	\WHILE{$\sum_{i \in L} x_{ij} < 1$ \AND $\beta < 1$}
		\FOR{every $i \in L$ in $\arg \max_{i'} \bigl \{ \frac{\partial C(\vect{x})}{\partial x_{i',j}} \bigr) \bigr \}$}
			\STATE \label{sub-step:allocation} Increase $x_{ij}$ by $dx$. 
			\STATE Increase $\beta$ and $\alpha_{j}$ by the following rules respectively,
			$$
			d\beta = \frac{\partial C(\vect{x})}{\partial x_{ij}} g(C(\vect{x})) dx
			\qquad \text{and} \qquad
			d\alpha_{j} = \frac{\partial C(\vect{x})}{\partial x_{ij}} \left(1 - \beta \right) dx.
			$$
		\ENDFOR
	\ENDWHILE
	\STATE For every $i$, assign $j$ to $i$ with probability $x_{ij}$.
\ENDFOR
\end{algorithmic}
\caption{Algorithm for Edge-Weighted Matching with Sub-Additive Rewards.}
\label{algo:sub-additive}
\end{algorithm}

Note that in the algorithm, $\frac{\partial C(\vect{x})}{\partial x_{ij}} = \E_{M'\sim \vect{x}_{-(i,j)}} [c(M' \cup (i,j)) - c(M')]$ can be computed 
(up to any precision) based on already arrival vertices (since $x_{(i',j')} = 0$ for every $j'$ unreleased so far).  

\paragraph{Primal/dual variables.}
Again, we define primal variables $z_{M} = \prod_{(i,j) \in M} x_{ij} \prod_{(i,j) \notin M} (1 - x_{ij})$
and dual variables $\alpha_{j}, \beta$ are constructed in the algorithm. 

\begin{lemma}
During the execution of the algorithm, the following invariant holds
$$
\beta = G(C(\vect{x}))
$$
\end{lemma}
\begin{proof}
Fix a vertex $i \in L$.
By the algorithm, whenever a variable $x_{ij}$ for some $j \in R$ is increased by an amound $dx$, 
$\frac{d\beta}{dx} = \frac{\partial C(\vect{x})}{\partial x_{ij}} g(C(\vect{x})) = \frac{\partial G(C(\vect{x}))}{\partial x_{ij}}$.
Together with the property $G(0) = G(C(\vect{0})) = 0$, the invariant follows.
\end{proof}

\begin{lemma}	\label{lem:sub-add-feasible}
For every $M \in \mathcal{M}$, it holds that
$$
\beta + \sum_{i \in L} \sum_{j \in M_{i}} \alpha_{j} \geq (1 - \kappa) c(M)
$$
In other words, the dual variables are feasible up to a factor $(1 - \kappa)$.
\end{lemma} 
\begin{proof}
Fix $M \in \mathcal{M}$. If $\beta \geq 1$ then the lemma inequality holds (recall that $c(M) \leq 1$). 
In the following, assume that $\beta < 1$. 
By the conditions of the loops, for every vertex $j \in M_{i}$,  
$\frac{d\alpha_{j}}{dx}$ must be at least 
$\frac{\partial C(\vect{x})}{\partial x_{ij}} \left(1 - \beta \right)$ since $\beta$ is non-decreasing. 
Besides, by the definition of multilinear extension, $C$ is linear function w.r.t $x_{ij}$. 
In other words, $\frac{\partial C(\vect{x})}{\partial x_{ij}}$ is constant when $x_{ij}$ varies.
Therefore,
$$
\alpha_{j} = \int_{0}^{1} \frac{d\alpha_{j}}{dx}dt 
\geq \int_{0}^{1} \frac{\partial C(\vect{x})}{\partial x_{ij}} \left(1 - \beta \right) dt 
= \frac{\partial C(\vect{x})}{\partial x_{ij}} \left(1 - \beta \right).
$$

We have
\begin{align*}
\beta + \sum_{i \in L} \sum_{j \in M_{i}} \alpha_{j} 
&\geq \beta + \sum_{i \in L} \sum_{j \in M_{i}} \frac{\partial C(\vect{x})}{\partial x_{ij}} \left(1 - \beta \right) \\
&\geq \beta + \left(1 - \beta \right) \cdot \min \biggl \{ 1,  \sum_{(i,j) \in M} \frac{\partial C(\vect{x})}{\partial x_{ij}} \biggr \}  \\
&= \min \biggl \{ 1,  \sum_{(i,j) \in M} \frac{\partial C(\vect{x})}{\partial x_{ij}} \biggr \} 
	+ \biggl( 1 - \min \biggl \{ 1,  \sum_{(i,j) \in M} \frac{\partial C(\vect{x})}{\partial x_{ij}} \biggr \} \biggr) \beta \\
&\geq \min \biggl \{ 1,  \sum_{(i,j) \in M} \frac{\partial C(\vect{x})}{\partial x_{ij}} \biggr \} \\
&= \min \biggl \{ 1, \sum_{(i,j) \in M} \E_{M' \sim \vect{x}_{-(i,j)}}[c(M' \cup (i,j)) - c(M')] \biggr \}  \\
&\geq \min \biggl \{ 1, \sum_{(i,j) \in M} \E_{M' \sim \vect{x}_{-(i,j)}}[ (1 - \kappa) c(i,j)] \biggr \}  \\
&= \min \biggl \{ 1,  (1 - \kappa) \sum_{(i,j) \in M}  c(i,j)  \biggr \} \\
&\geq (1 - \kappa) c(M).
\end{align*}
The second inequality holds since $1 - \beta \geq 0$. 
The second equality is due to property (\ref{eq:increase-C}) of multilinear extensions. 
The third inequality follows the definition of curvature. 
The last inequality is due to  the sub-additivity of $c$ and $c(M) \leq 1$. 
The lemma inequality follows.
\end{proof}

\begin{theorem}
Algorithm~\ref{algo:sub-additive} is $(1 - \kappa)(1 - 1/e)$-competitive.
\end{theorem}
\begin{proof}
We bound the increasing rates of the primal and the dual. 
Assume that $dx$ amount of $j$ is allocated to $i$. The increase in the primal is $\frac{\partial C(\vect{x})}{\partial x_{ij}} dx$.
The increase of the dual is 
\begin{align*}
d\alpha_{j} + d\beta_{i} &= \frac{\partial C(\vect{x})}{\partial x_{ij}} \left(1 - \beta \right) dx + \frac{\partial C(\vect{x})}{\partial x_{ij}} g(C(\vect{x})) dx \\
&=   \biggl( 1 - G(C(\vect{x}))  + g(C(\vect{x})) \biggr) \frac{\partial C(\vect{x})}{\partial x_{ij}} dx
=   \frac{1}{1 -1/e} \cdot \frac{\partial C(\vect{x})}{\partial x_{ij}} dx.
\end{align*}
By Lemma~\ref{lem:sub-add-feasible} and the weak duality, the competitive ratio follows.
\end{proof}



\section{Online Fractional Weighted Matching with Concave Rewards}	\label{sec:concave}
The main goal of this section is to build the connection with the primal-dual scheme in
\cite{DevanurJain12:Online-matching} by showing that the latter can be described and analyzed similarly as algorithms in previous sections.  
Consider the online weighted matching problem in the general model 
in which the multilinear extension $C: [0,1]^{|E|} \rightarrow [0,1]$ (of the reward function $c$) is \emph{concave}.

%

Given a function $\vect{v}: [0,1]^{|E|} \rightarrow [0,1]^{|E|}$, for $\vect{x} \in [0,1]^{|E|}$, 
we can write $\vect{v}(\vect{x}) = (v_{i,j}(\vect{x}))_{(i,j) \in E}$.
Define $r > 0$ as the largest constant such that the following system of differential equations has a solution $\vect{v}: [0,1]^{|E|} \rightarrow [0,1]^{|E|}$
\begin{align}	
\frac{1}{r} \cdot \frac{\partial C(\vect{x})}{\partial x_{ij}}
&= \frac{\partial C(\vect{v}(\vect{x}))}{\partial v_{ij}(\vect{x})} 
- \biggl \langle \vect{v}(\vect{x}), \frac{\partial}{\partial x_{ij}} \nabla_{\vect{v}} C(\vect{v}(\vect{x})) \biggr \rangle \qquad \forall i \in L, j \in R, 	\label{eq:diff-1} \\
\frac{\partial \vect{v}(\vect{x})}{\partial x_{ij}} &\geq 0	\qquad \forall i \in L, j \in R, 	\label{eq:diff-2} \\
\frac{\partial^{2} C(\vect{v}(\vect{x}))}{\partial x_{ij} \partial v_{ij}(\vect{x})} &\leq 0 \qquad \forall i \in L, j \in R,	\label{eq:diff-3}
\end{align}
with boundary conditions $v(\vect{0}) = \vect{0}$. Note that the last inequality guarantees that 
$\frac{\partial C(\vect{v}(\vect{x}))}{\partial v_{ij}(\vect{x})}$ is non-increasing when $x_{ij}$ increases.

Let $\vect{v}$ be a solution of the system with $r$. 
Denote 
$\nabla_{\vect{v}} C(\vect{v}(\vect{x})) = \bigl(  \frac{\partial C(\vect{v}(\vect{x}))}{\partial v_{ij}(\vect{x})} \bigr)_{(i,j) \in E}$.
Consider the following algorithm, building from the salient ideas of the previous ones. 

\begin{algorithm}[ht]
\begin{algorithmic}[1]  
\STATE All primal and dual variables are initially set to 0.
\FOR{each arrival of a new vertex $j$}
	\WHILE{$\sum_{i \in L} x_{ij} < 1$}
		\FOR{every $i \in L$ in $\arg \max_{i'} \bigl \{  \frac{\partial C(\vect{v}(\vect{x}))}{\partial v_{i',j}(\vect{x})} \bigr \}$}
			\STATE \label{sub-step:allocation} Allocate a $dx = dx_{ij}$ amount of $j$ to $i$. 
			\STATE Increase $\alpha_{j}$ by the following rule:
			$$
			\frac{d\alpha_{j}}{dx_{ij}} = \frac{\partial C(\vect{v}(\vect{x}))}{\partial v_{ij}(\vect{x})}.
			$$
		\ENDFOR
		\STATE Always maintain $\beta = C(\vect{v}(\vect{x})) - \bigl \langle \vect{v}(\vect{x}), \nabla_{\vect{v}} C(\vect{v}(\vect{x})) \bigr \rangle$.
	\ENDWHILE
\ENDFOR
\end{algorithmic}
\caption{Algorithm for Edge-Weighted Matching with Concave Rewards.}
\label{algo:concave}
\end{algorithm}

\begin{lemma}
For every $M \in \mathcal{M}$, it holds that
$$
\beta + \sum_{i \in L} \sum_{j \in M_{i}} \alpha_{j} \geq c(M)
$$
\end{lemma} 
\begin{proof}
Fix $M \in \mathcal{M}$.
For any vertex $i \in L$ and $j \in M_{i}$,
at every time during the execution of the algorithm, 
the increasing rate of $\alpha_{j}$ is at least 
the current value of $\frac{\partial C(\vect{v}(\vect{x}))}{\partial v_{ij}(\vect{x})}$. 
Moreover, the latter is non-increasing (by (\ref{eq:diff-3})).  
Therefore,
$$
\alpha_{j} = \int_{0}^{1} \frac{\partial C(\vect{v}(\vect{x}))}{\partial v_{ij}(\vect{x})}dt 
\geq \frac{\partial C(\vect{v}(\vect{x}))}{\partial v_{ij}(\vect{x})}.
$$

We have 
\begin{align*}
\beta + \sum_{i \in L} \sum_{j \in M_{i}} \alpha_{j} 
&\geq C(\vect{v}(\vect{x})) - \bigl \langle \vect{v}(\vect{x}), \nabla_{\vect{v}} C(\vect{v}(\vect{x})) \bigr \rangle + \sum_{(i,j) \in M} \frac{\partial C(\vect{v}(\vect{x}))}{\partial v_{ij}(\vect{x})} \\
&= C(\vect{v}(\vect{x})) + \bigl \langle \nabla_{\vect{v}} C(\vect{v}(\vect{x})), \one_{M} - \vect{v}(\vect{x}) \bigr \rangle \\
&\geq C(\vect{v}(\vect{x})) + C(\one_{M}) - C(\vect{v}(\vect{x})) \\
&= C(\one_{M}) = c(M)
\end{align*}
where $\one_{M}$ is the indicator vector of $M$, i.e., $(\one_{M})_{e} = 1$ if $e \in M$ and $(\one_{M})_{e} = 0$ otherwise. 
The last inequality is due to the concavity of $C$.
\end{proof}

\begin{theorem}	\label{thm:concave}
Algorithm~\ref{algo:concave} is $r$-competitive for the fractional online matching problem under the assumption that 
$C$ is concave. 
\end{theorem}
\begin{proof}
Initially, when no online vertex is released, the objective values of the primal and the dual are 0. 
At any time in the execution of the algorithm when an online vertex $j$ is released,
the increasing rate of the dual is 
\begin{align*}
\frac{d\alpha_{j}}{dx_{ij}} + \frac{d\beta}{dx_{ij}} 
&= \frac{\partial C(\vect{v}(\vect{x}))}{\partial v_{ij}(\vect{x})} 
- \frac{\partial}{\partial x_{ij}} \biggl( C(\vect{v}(\vect{x})) - \bigl \langle \vect{v}(\vect{x}), \nabla_{\vect{v}} C(\vect{v}(\vect{x})) \bigr \rangle \biggr) \\
&= \frac{\partial C(\vect{v}(\vect{x}))}{\partial v_{ij}(\vect{x})} 
- \biggl \langle \vect{v}(\vect{x}), \frac{\partial}{\partial x_{ij}} \nabla_{\vect{v}} C(\vect{v}(\vect{x})) \biggr \rangle.
\end{align*}
Besides, the increasing rate of the primal is $\frac{\partial C(\vect{x})}{\partial x_{ij}}$.
The competitive ratio follows the definition of $r$.
\end{proof}

\paragraph{Connection to the scheme of \citet{DevanurJain12:Online-matching}.}
Consider the setting in \cite{DevanurJain12:Online-matching} in which each edge between $i \in L$ and $j \in R$ has weight $w_{ij} \geq 0$ 
and the reward function of each offline vertex $i \in L$ can be expressed as $P(\sum_{j} w_{ij}x_{ij})$ 
where $P: \mathbb{R}^{+} \rightarrow \mathbb{R}^{+}$ is a 1-dim concave function and 
$x_{ij}$ is the fraction of online vertex $j$ assigned to $i$. 
The total reward is $\sum_{i \in L} P(\sum_{j} w_{ij}x_{ij})$.
In this setting, using $\sum_{i \in L} P$ instead of $C$ in (\ref{eq:diff-1}, \ref{eq:diff-2}, \ref{eq:diff-3}) and Algorithm~\ref{algo:concave}, 
after a simplification (due to the separability of the reward on each offline vertex in the total reward), 
the system of differential equations characterizing the competitive ratio reads
\begin{align*}
\frac{1}{r} \cdot P'(u)
&= P'(v(u)) - [v(u) P''(v(u))]\frac{dv(u)}{du} \\
\frac{dv(u)}{du} &\geq 0
\end{align*}
with boundary condition $v(0) = 0$. 
The system becomes simpler as $P$ is one-variable function. Note that in this system, we name (1-dim) variable $u$ (instead of $x$)
and the condition (\ref{eq:diff-3}), written as
$$
d \left( \frac{\partial P(v(u))}{\partial v(u)} \right)/du = P''(v(u)) \frac{dv(u)}{du} \leq 0,
$$ 
always holds (by the concavity of $P$ and the condition $\frac{dv(u)}{du} \geq 0$).
This system is exactly the one given in \cite{DevanurJain12:Online-matching}. For completeness, we give it below.
\begin{align*}
\frac{1}{r} \cdot P'(u)
&= P'(v(u)) + Y'(v(u))\frac{dv(u)}{du} \\
\frac{dv(u)}{du} &\geq 0
\end{align*}
with boundary condition $Y(0) = 0$ where the function $Y(v) = P(v) - vP'(v)$.

\section{Conclusion}
In this paper, we have presented a primal-dual framework with configuration LPs for the online bipartite matching problem 
that unifies previous approaches and provides optimal $(1-1/e)$ competitive ratio in some models, resolving long-standing open questions.
We believe that the use of primal-dual with configuration LPs (for example, to circumvent hard constraints, etc) 
will find other applications and lead to improvements in different problems. Related to our paper, an open question is to study whether the bounds 
given in Sections~\ref{sec:sub-additive} and \ref{sec:concave} are tight.

\bibliographystyle{plainnat}
\bibliography{matching} 

\begin{thebibliography}{26}
\providecommand{\natexlab}[1]{#1}
\providecommand{\url}[1]{\texttt{#1}}
\expandafter\ifx\csname urlstyle\endcsname\relax
  \providecommand{\doi}[1]{doi: #1}\else
  \providecommand{\doi}{doi: \begingroup \urlstyle{rm}\Url}\fi

\bibitem[Aggarwal et~al.(2011)Aggarwal, Goel, Karande, and
  Mehta]{AggarwalGoel11:Online-vertex-weighted}
Gagan Aggarwal, Gagan Goel, Chinmay Karande, and Aranyak Mehta.
\newblock Online vertex-weighted bipartite matching and single-bid budgeted
  allocations.
\newblock In \emph{Proc. 22nd Symposium on Discrete Algorithms}, pages
  1253--1264, 2011.

\bibitem[Balcan and Harvey(2012)]{BalcanHarvey12:Learning-Submodular}
Maria{-}Florina Balcan and Nicholas J.~A. Harvey.
\newblock Learning submodular functions.
\newblock In \emph{Proc. Machine Learning and Knowledge Discovery in
  Databases}, pages 846--849, 2012.

\bibitem[Birnbaum and Mathieu(2008)]{BirnbaumMathieu08:On-line-bipartite}
Benjamin Birnbaum and Claire Mathieu.
\newblock On-line bipartite matching made simple.
\newblock \emph{Acm Sigact News}, 39\penalty0 (1):\penalty0 80--87, 2008.

\bibitem[Buchbinder et~al.(2007)Buchbinder, Jain, and
  Naor]{BuchbinderJain07:Online-primal-dual}
Niv Buchbinder, Kamal Jain, and Joseph~Seffi Naor.
\newblock Online primal-dual algorithms for maximizing ad-auctions revenue.
\newblock In \emph{European Symposium on Algorithms}, pages 253--264, 2007.

\bibitem[Conforti and
  Cornu{\'e}jols(1984)]{ConfortiCornuejols84:Submodular-set-functions}
Michele Conforti and G{\'e}rard Cornu{\'e}jols.
\newblock Submodular set functions, matroids and the greedy algorithm: tight
  worst-case bounds and some generalizations of the rado-edmonds theorem.
\newblock \emph{Discrete applied mathematics}, 7\penalty0 (3):\penalty0
  251--274, 1984.

\bibitem[Devanur and Jain(2012)]{DevanurJain12:Online-matching}
Nikhil~R Devanur and Kamal Jain.
\newblock Online matching with concave returns.
\newblock In \emph{Proc. 44th ACM symposium on Theory of computing}, pages
  137--144, 2012.

\bibitem[Devanur et~al.(2013)Devanur, Jain, and
  Kleinberg]{DevanurJain13:Randomized-primal-dual}
Nikhil~R Devanur, Kamal Jain, and Robert~D Kleinberg.
\newblock Randomized primal-dual analysis of ranking for online bipartite
  matching.
\newblock In \emph{Proc. 24th Symposium on Discrete algorithms}, pages
  101--107, 2013.

\bibitem[Devanur et~al.(2016)Devanur, Huang, Korula, Mirrokni, and
  Yan]{DevanurHuang16:Whole-page-optimization}
Nikhil~R Devanur, Zhiyi Huang, Nitish Korula, Vahab~S Mirrokni, and Qiqi Yan.
\newblock Whole-page optimization and submodular welfare maximization with
  online bidders.
\newblock \emph{ACM Transactions on Economics and Computation (TEAC)},
  4\penalty0 (3):\penalty0 1--20, 2016.

\bibitem[D{\"u}rr and Thang(2021)]{DurrThang21:Implementation-on-upper}
Christoph D{\"u}rr and Nguyen~Kim Thang.
\newblock Implementation on upper triangular graph, September 2021.
\newblock URL \url{https://www.ibisc.univ-evry.fr/~thang/matching.html}.

\bibitem[Fahrbach et~al.(2020)Fahrbach, Huang, Tao, and
  Zadimoghaddam]{FahrbachHuang20:Edge-weighted-online}
Matthew Fahrbach, Zhiyi Huang, Runzhou Tao, and Morteza Zadimoghaddam.
\newblock Edge-weighted online bipartite matching.
\newblock In \emph{Proc. 61st Symposium on Foundations of Computer Science},
  2020.

\bibitem[Feldman et~al.(2009)Feldman, Korula, Mirrokni, Muthukrishnan, and
  P{\'a}l]{FeldmanKorula09:Online-ad-assignment}
Jon Feldman, Nitish Korula, Vahab Mirrokni, Shanmugavelayutham Muthukrishnan,
  and Martin P{\'a}l.
\newblock Online ad assignment with free disposal.
\newblock In \emph{Workshop on internet and network economics (WINE)}, pages
  374--385, 2009.

\bibitem[Goemans et~al.(2009)Goemans, Harvey, Iwata, and
  Mirrokni]{GoemansHarvey09:Approximating-submodular}
Michel~X Goemans, Nicholas~JA Harvey, Satoru Iwata, and Vahab Mirrokni.
\newblock Approximating submodular functions everywhere.
\newblock In \emph{Proc. 20th Symposium on Discrete algorithms}, pages
  535--544, 2009.

\bibitem[Goyal and Udwani(2020)]{GoyalUdwani20:Online-Matching}
Vineet Goyal and Rajan Udwani.
\newblock Online matching with stochastic rewards: Optimal competitive ratio
  via path based formulation.
\newblock In \emph{Proc. 21st ACM Conference on Economics and Computation},
  pages 791--791, 2020.

\bibitem[Huang and Zhang(2020)]{HuangZhang20:Online-primal}
Zhiyi Huang and Qiankun Zhang.
\newblock Online primal dual meets online matching with stochastic rewards:
  configuration lp to the rescue.
\newblock In \emph{Proc. 52nd ACM Symposium on Theory of Computing}, pages
  1153--1164, 2020.

\bibitem[Huang et~al.(2020{\natexlab{a}})Huang, Kang, Tang, Wu, Zhang, and
  Zhu]{HuangKang20:Fully-online}
Zhiyi Huang, Ning Kang, Zhihao~Gavin Tang, Xiaowei Wu, Yuhao Zhang, and Xue
  Zhu.
\newblock Fully online matching.
\newblock \emph{Journal of the ACM (JACM)}, 67\penalty0 (3):\penalty0 1--25,
  2020{\natexlab{a}}.

\bibitem[Huang et~al.(2020{\natexlab{b}})Huang, Zhang, and
  Zhang]{HuangZhang20:Adwords-in-a-Panorama}
Zhiyi Huang, Qiankun Zhang, and Yuhao Zhang.
\newblock Adwords in a panorama.
\newblock In \emph{Proc. 61st Symposium on Foundations of Computer Science},
  2020{\natexlab{b}}.

\bibitem[Iyer et~al.(2013)Iyer, Jegelka, and
  Bilmes]{IyerJegelka13:Curvature-and-optimal}
Rishabh~K Iyer, Stefanie Jegelka, and Jeff~A Bilmes.
\newblock Curvature and optimal algorithms for learning and minimizing
  submodular functions.
\newblock In \emph{Advances in Neural Information Processing Systems}, pages
  2742--2750, 2013.

\bibitem[Kapralov et~al.(2013)Kapralov, Post, and
  Vondr{\'a}k]{KapralovPost13:Online-submodular}
Michael Kapralov, Ian Post, and Jan Vondr{\'a}k.
\newblock Online submodular welfare maximization: Greedy is optimal.
\newblock In \emph{Proc. 24th ACM-SIAM Symposium on Discrete algorithms}, pages
  1216--1225, 2013.

\bibitem[Karp et~al.(1990)Karp, Vazirani, and
  Vazirani]{KarpVazirani90:An-optimal-algorithm}
Richard~M Karp, Umesh~V Vazirani, and Vijay~V Vazirani.
\newblock An optimal algorithm for on-line bipartite matching.
\newblock In \emph{Proc. 22nd ACM symposium on Theory of computing}, pages
  352--358, 1990.

\bibitem[Korula et~al.(2013)Korula, Mirrokni, and
  Zadimoghaddam]{KorulaMirrokni13:Bicriteria-online}
Nitish Korula, Vahab~S Mirrokni, and Morteza Zadimoghaddam.
\newblock Bicriteria online matching: Maximizing weight and cardinality.
\newblock In \emph{International conference on web and internet economics},
  pages 305--318, 2013.

\bibitem[Mehta(2013)]{Mehta13:Online-Matching}
Aranyak Mehta.
\newblock Online matching and ad allocation.
\newblock \emph{Foundations and Trends{\textregistered} in Theoretical Computer
  Science}, 8\penalty0 (4):\penalty0 265--368, 2013.

\bibitem[Mehta and Panigrahi(2012)]{MehtaPanigrahi12:Online-matching}
Aranyak Mehta and Debmalya Panigrahi.
\newblock Online matching with stochastic rewards.
\newblock In \emph{53rd Symposium on Foundations of Computer Science}, pages
  728--737, 2012.

\bibitem[Mehta et~al.(2007)Mehta, Saberi, Vazirani, and
  Vazirani]{MehtaSaberi07:Adwords-and-generalized}
Aranyak Mehta, Amin Saberi, Umesh Vazirani, and Vijay Vazirani.
\newblock Adwords and generalized online matching.
\newblock \emph{Journal of the ACM}, 54\penalty0 (5):\penalty0 22--es, 2007.

\bibitem[Sviridenko et~al.(2017)Sviridenko, Vondr{\'a}k, and
  Ward]{SviridenkoVondrak17:Optimal-approximation}
Maxim Sviridenko, Jan Vondr{\'a}k, and Justin Ward.
\newblock Optimal approximation for submodular and supermodular optimization
  with bounded curvature.
\newblock \emph{Mathematics of Operations Research}, 2017.

\bibitem[Thang(2020)]{Thang20:Online-Primal-Dual}
Nguyen~Kim Thang.
\newblock Online primal-dual algorithms with configuration linear programs.
\newblock In \emph{Proc. 31st International Symposium on Algorithms and
  Computation}, 2020.

\bibitem[Vondr{\'a}k(2010)]{Vondrak10:Submodularity-and-Curvature:}
Jan Vondr{\'a}k.
\newblock Submodularity and curvature: The optimal algorithm.
\newblock \emph{RIMS Kokyuroku Bessatsu}, 2010.

\end{thebibliography}

\end{document}